\documentclass[aps,nofootinbib,showkeys
,showpacs]{revtex4} 
\usepackage{subfig}
\usepackage{float}
\usepackage{epsfig}
\usepackage{hyperref}
\usepackage[usenames,dvipsnames]{color}
\usepackage{amssymb}
\usepackage{amsmath}
\usepackage{latexsym}
\usepackage{amsfonts}
\usepackage{bm}
\newcommand{\be}{\begin{equation}}
\newcommand{\ee}{\end{equation}}
\newcommand{\bea}{\begin{eqnarray}}
\newcommand{\eea}{\end{eqnarray}}

\newcommand{\bwt}{ \begin{widetext}}
\newcommand{\ewt}{ \end{widetext}}

\newcommand{\beq}{\begin{equation}}
\newcommand{\eeq}{\end{equation}}

\begin{document}
\title{Constraining $f(R)$ theories with cosmography}
\author{\ Florencia Anabella Teppa Pannia}
\email{fteppa@fcaglp.unlp.edu.ar}
\affiliation{Facultad de Ciencias Astron\'omicas y Geof\'{\i}sicas, Universidad Nacional de La Plata, Paseo del Bosque s/n, B1900FWA, Argentina.
}
\author{\ Santiago Esteban Perez Bergliaffa}
\email{sepbergliaffa@gmail.com}
\affiliation{Departamento de F\'{\i}sica Te\'{o}rica,
Instituto de F\'{\i}sica, Universidade do Estado de Rio de
Janeiro, CEP 20550-013, Rio de Janeiro, Brazil.}
\vspace{.5cm}

\begin{abstract}

A method to set constraints on the parameters of 
extended theories of gravitation 
is presented. It is based on the comparison of two series expansions 
of any observable that depends on $H(z)$. 
The first expansion is of the
cosmographical type, 
while the second uses the dependence of $H$
with $z$ furnished by 
a given type of extended theory.
When applied to $f(R)$ theories together with 
the redshift drift, the method yields limits 
on the parameters of two examples (the theory of Hu and Sawicki \cite{HuSawicki2007}, and the 
exponential gravity introduced by Linder \cite{Linder2009})
that are compatible with 
or more stringent
than the existing ones, as well as a 
limit for a previously unconstrained parameter.

\end{abstract}

\maketitle


\section{Introduction}
\label{Intro}

The interpretation of several sets of data (such as those obtained from
type Ia supernovae, large scale structure, baryon acoustic oscillations, and the cosmic microwave background)
in the framework of the Standard Cosmological Model (SCM) (based on General Relativity (GR) and the Cosmological Principle)  
indicates 
that the universe is currently undergoing a phase of accelerated expansion. The
most commonly accepted candidates to source such an expansion (namely, the cosmological constant, and some unknown type of matter dubbed ``dark energy''
\footnote{For a complete list of candidates see \cite{Li2011}.}) 
are not free of problems. While the energy density
associated with the cosmological constant that is inferred
from astronomical observations is approximately 120 orders
of magnitude lower than the value predicted by field theory (see for instance \cite{Capozziello2011}),
the scalar field used to model dark energy has features that are alien to those displayed by the scalar fields of particle physics \cite{Sotiriou2010}. 
An alternative way to describe the accelerated expansion
is to assume that it is produced by the dynamics of a theory which differs from GR 
after matter domination.
Among these, the so-called $f(R)$ theories, with action given by 
$$
S = \int \sqrt{-g} f(R) {\rm d}^4 x,
$$
are the simplest
generalization of the Einstein-Hilbert Lagrangian.
The dependence of the function $f$ on the scalar curvature $R$ is to be  determined by several criteria (such as matter stability \cite{Dolgov2003}, 
absence of ghost modes in the cosmological perturbations \cite{Carroll2003},
correct succession of cosmological eras \cite{Amendola2007}, and the stability of cosmological perturbations \cite{Bean2007})\footnote{For reviews about different aspects of $f(R)$ theories, see
\citep{Sotiriou2010,deFelice2010,Capozziello2011,Nojiri2011}.} .
Several forms for $f(R)$ 
have been constructed 
in order to successfully satisfy 
these constraints (for instance those given in \citep{Starobinski2007,HuSawicki2007}),
allowing in principle (potentially small) deviations from GR,
quantified by some of the parameters of the $f(R)$.
We shall introduce here a method that can be used to set limits on the parameters of a given $f(R)$, which is based in the comparison of two series expansions in terms of the redshift $z$. The first one is that of any observable quantity given in terms of $H(z)$ 
, and the second, the corresponding cosmographic 
expansion. While the former depends on the dynamics of the theory, the latter does not (see Sect.~\ref{sec:cosmography})\footnote{The cosmographical approach has already led to 
interesting results  
in the framework of $f(R)$ theories, see
\citep{Capozziello2008,Capozziello2011b,Aviles2012,Shafieloo2012}.}.
The order-by-order comparison of these expansions yields relations among $f$, its derivatives w.r.t. $R$, and the kinematical parameters (some of which are determined by observation), all at $t=\ $ today. By rewritting these relations in terms of the parameters of a given $f$, the abovementioned limits can be obtained.

It is important to emphasize that the method does not rely on the actual measurement of the observable: it only demands that the expression of the observable obtained using the dynamics coincides with that obtained in a dynamic-independent way (namely, using cosmography). 
Although it can be applied to any observable expressed in terms of $H(z)$, yielding for each observable different limits on the parameters of the $f(R)$,
the method is more useful in the case of yet-to-be measured quantities, of which 
only the cosmographical form has been determined.

Currently a lot of effort is devoted to the study of quantities and effects that have the potential of discriminating between different $f(R)$ models, and between these models and GR. Among them, we can mention the growth rate of matter density perturbations (see for instance 
\cite{Carroll2006}), 
the enhanced brightness of dwarf galaxies \cite{Davies2012}, the modifications of the 21cm power spectrum at reionisation \cite{Brax2012}, 
the specific angular momentum of galactic halos \cite{Lee2012}, and
the number counts of peaks in weak
lensing maps \cite{Cardone2012}. 
We shall apply our method to the Redshift Drift (RD), that is, the time variation of the cosmological redshift caused by the expansion of the Universe. 
The RD was first considered by \citet{Sandage1962}, and the effects of a nonzero cosmological constant on it were presented in 
\cite{McVittie1962}. 
As discussed in 
\cite{Quercellini2010}, 
its measurement is feasible in the near future.
As soon as data related to the RD become available, they could be compared with the prediction of a given $f(R)$, adjusting the parameters of the theory to describe the data. We propose here the alternative route presented above, namely the comparison of the 
``cosmographical RD'' with the ``dynamical RD'', the results of which must be compatible with those that will come from the actual measurements.\footnote{Note that are several effects (such as those coming from
the peculiar acceleration in nearby clusters and galaxies, and the
peculiar velocity of the source) that should be taken into account when comparing a theoretical prediction for the RD 
with observations. This is not the case for the method proposed here.}

When compared to other cosmological observables, the RD has the advantage that it directly 
tests the dependence of the Hubble parameter with the redshift, hence
probing the dynamics of the scale factor. 
Another feature of  this observable  
is that it does not depend on details of the source (such as the 
absolute luminosity), or on the definition of a standard ruler. 
As demonstrated in \cite{Uzan2008},
the RD  would allow the test of the Copernican Principle, thus checking for any degree of radial inhomogeneity. This issue was further discussed in \citep{QuartinAmendola2010}, where it was shown that  
the RD is positive 
for sources with $z<2$ in the $\Lambda$CDM, while in Lem\^aitre-Tolman-Bondi models\footnote{See for instance \citep{Plebanski2006}.}
is negative for sources observed from the symmetry center \citep{Yoo2011}.
The RD can also be used to 
constrain phenomenological parametrizations of dynamical dark energy
models (see \cite{Vielzeuf2012,Martinelli2012,Moraes2011}). 
To use the RD as an example of our method,
we shall work out in Sect.~\ref{sec:cosmography} the series expansion of this observable in terms of the cosmological redshift $z$ and the time derivatives of the scale factor ({\it i.e.} the kinematical parameters).
Since the RD depends on the explicit form of the Hubble parameter $H(z)$, we shall show in Sect.~\ref{sec:fR} that its series expansion in $z$ can be written in terms of $f(R)$ and its derivatives, using the equations of motion (EoM) of a yet unspecified $f(R)$ theory in the metric version. 
By comparing the two series, it will be shown that there exists relations between $f(R)$, its derivatives, and 
the kinematical parameters, which impose constraints on the parameters of a given $f(R)$. These constraints will be analyzed on two examples: that proposed by \citet{HuSawicki2007} (see Sect.\ref{hs}),   
and the exponential gravity theory introduced by \citet{Linder2009}(see Sect.\ref{lin}). In both cases, we find limits 
on the parameters of these $f(R)$ theories 
that are compatible with 
or more stringent
than the existing ones, as well as a 
limit for a previously unconstrained parameter.
We close in Sect.~\ref{sec:discussion} with some remarks.

\section{A cosmographical approach to the redshift drift}
\label{sec:cosmography}

Cosmography is a mathematical framework for the description of the universe, based entirely on the Cosmological Principle, 
 and on those parts of  GR  
that follow directly from the Principle of Equivalence \citep{Weinberg1972}. It is inherently kinematic, in the sense that it is independent of the dynamics obeyed by the scale factor $a(t)$. 
In this section we shall  
present the calculation, in the context of cosmography, that leads to the series expansion of the RD in terms of $z$, assuming only that spacetime is homogeneous and isotropic. 

The redshift of a photon emitted by a source at time $t$ that reaches the observer at time $t_{obs}$ is given by
\be
\label{z1}
z(t)=\frac{a(t_{obs})}{a(t)}-1. 
\ee
The time variation of the redshift is obtained by comparing this expression with the one corresponding to a photon emitted at $t'=t+\Delta t$,  that is $\Delta z=z(t+\Delta t) - z(t)$. 
To first order in $\Delta t_{obs}$ and $\Delta t$, it follows that \citep{Loeb1998}
\be
\label{RD1}
\frac{\Delta z}{\Delta t_{obs}} = \left[\frac{\dot a(t_{obs})-\dot a(t)}{a(t)}\right].
\ee
Using the definition of $z$ and $H(z)$ in this equation we get 
the expression of the RD in terms of $z$, namely
\be
\label{RDd}
\frac{\Delta z}{\Delta t_{obs}} = (1+z)H_{obs}-H(z).
\ee
Next, an expansion of $H$ in powers of $z$ will be obtained, using the cosmographical approach, while in the next section we will exhibit
the analogous expansion using the form of $H$ determined by a given $f(R)$ theory. 
The series development of the scale factor around $t_0=t_{obs}$ is given by
\be
\label{factor_a}
a(t)=a_0 \left[1+ H_0(t-t_0)-\frac{1}{2} q_0 H_0^2 (t-t_0)^2 
+\frac{1}{3!} j_0 H_0^3 (t-t_0)^3
+ \frac{1}{4!} s_0 H_0^4 (t-t_0)^4 + {\cal O}([t-t_0]^5)\right],
\ee
where the so-called kinematical parameters are defined by
$$
H(t)\equiv+\frac{1}{a(t)}\frac{{\rm d}a}{{\rm d}t},\ \ q(t)\equiv-\frac{1}{a(t)} \frac{{\rm d}^2a}{{\rm d}t^2} \left[\frac{1}{a(t)} \frac{{\rm d}a}{{\rm d}t} \right]^{-2},\ \ j(t)\equiv +\frac{1}{a(t)} \frac{{\rm d}^3a}{{\rm d}t^3} \left[\frac{1}{a(t)} \frac{{\rm d}a}{{\rm d}t} \right]^{-3},\ 
s(t)\equiv +\frac{1}{a(t)} \frac{{\rm d}^4a}{{\rm d}t^4} \left[\frac{1}{a(t)} \frac{{\rm d}a}{{\rm d}t} \right]^{-4}.
$$
In order to use Eq.~(\ref{factor_a}) for the calculation of the RD, we need to express $t$ in terms of known quantities. This can be achieved through the physical distance travelled by a photon emitted at $t$ and observed at $t_0$, given by
\be
\label{D}
D=c\int {\rm d}t= c(t_0-t).
\ee
A relation between $D$ and $z$ can be obtained 
from Eq.~(\ref{z1}) \cite{Visser2005}:
\be
\label{z2}
1+z=\frac{a(t_0)}{a(t)}=\frac{a(t_0)}{a(t_0-D/c)}.
\ee
Performing a Taylor series expansion, Eq.~(\ref{z2}) yields 
\be
z(D)=\frac{H_0D}{c} + \frac{2+q_0}{2}\frac{H_0^2D^2}{c^2}+ \frac{6(1+q_0)+j_0}{6}\frac{H_0^3D^3}{c^3}+  {\cal O}\left(\left[\frac{H_0D}{c}\right]^4\right),
\ee
which can be inverted to
{\small{
\be
\label{D(z)}
D(z)=\frac{c\ z}{H_0} \left[ 1-\left(1+\frac{q_0}{2}\right)z + \left(1+q_0 +\frac{q_0^2}{2}- \frac{j_0}{6}\right) z^2 - \left(1+ \frac{3}{2} q_0(1+q_0) + \frac{5}{8} q_0^3- \frac{1}{2} j_0 -\frac{5}{12} q_0j_0-\frac{s_0}{24}\right) z^3 + {\cal O}(z^4) \right].
\ee
}} \\

Setting $t=t_0-D/c$, the Taylor expansion of expression (\ref{RD1}) yields
\be
\frac{\Delta z}{\Delta t_0}(D)=-q_0H_0^2\frac{D}{c}-\frac{1}{2}H_0^3(j_0+2q_0)\left(\frac{D}{c}\right)^2-\frac{1}{6}H_0^4[3q_0(q_0+2)-s_0+3j_0]\left(\frac{D}{c}\right)^3+ {\cal O}(D^4).
\ee
Lastly, using Eq.~(\ref{D(z)}) we can write the RD as a power series in $z$, with coefficients that are functions of the kinematical parameters in the form
\be
\label{RDz3}
\frac{\Delta z}{\Delta t_0}(z)= -H_0 q_0 z
+\frac{1}{2}H_0\left( q_0^2-j_0 \right) z^2
+\frac{1}{2}H_0\left[\frac{1}{3}\left(s_0+4q_0j_0\right)+j_0-q_0^2-q_0^3\right] z^3
+ {\cal O}(z^4). 
\ee  
This equation gives the cosmographical expression of the RD up to the third order in the redshift of the source, in terms of the value of the kinematical parameters at the present epoch (whose values are known from observation, see Sect.\ref{hs}). 
Let us remark that Eq.~(\ref{RDz3}) is completely independent of the dynamics
obeyed by the gravitational field. Hence, any viable theory must yield a prediction for the RD compatible with it.
In the next section, this expression 
will be compared with that 
obtained using the dynamics of an arbitrary $f(R)$ theory.

\section{The Redshift Drift in $f(R)$ Theories}
\label{sec:fR}
 
Let us recall that the RD can be expressed in terms of $H(z)$ as follows
\be
\label{RDd2}
\frac{\Delta z}{\Delta t_0} = (1+z)H_0-H(z).
\ee
In the case of the SCM, 
the RD can be written as a function of the cosmological parameters $H_0$, $\Omega_{m,0}$, $\Omega_{r,0}$ and $\Omega_{\Lambda,0}$ 
using the exact expression for $H(z)$ 
as follows \citep{Loeb1998}:
\be
\frac{ \Delta z}{ \Delta t_0}=H_0 \left[(1+z)-\left( \Omega_{m,0}(1+z)^3+
\Omega_{r,0}(1+z)^4+ \Omega_{\Lambda,0}\right)^{1/2}\right],
\label{rdscm}
\ee
where the subindex 0 means that the corresponding quantity is evaluated at
$t = $ today. 
In the case of $f(R)$ theories, the expression for $H$ that follows from the
variation of the action 
\be
S=\int {\rm d}^4x \sqrt{-g} \left[{f(R)}+ {\cal L}_{\rm matter}\right] \nonumber
\ee
w.r.t. the metric must be used. For the FLRW metric and considering a pressureless cosmological fluid, these equations are (see for instance \cite{Kerner1982})
\bea 
\label{EOM1}
f'R -2f-3f''\left(\ddot R + \frac{3\dot a \dot R}{a} \right) - 3f'''\dot R^2 =T,& \\
\label{EOM2}
& \nonumber \\
f' R_{tt} + \frac{1}{2} f + 3f''\frac{\dot a \dot R}{a} =T_{tt},&
\eea
where dot and prime denote, respectively, derivative w.r.t. $t$ and $R$, $R_{tt}=3\ddot{a}/a$, $R=6(\frac{\ddot a}{a}+ \frac{\dot a^2}{a^2})$, and $T$ is the trace of the energy-momentum tensor.
From these, the following  relation can be obtained \citep{Capozziello2008}: 
\be
H=\frac{1}{6\dot R f''}\left(6H^2f'-2\rho_m-f+Rf'\right),
\label{hcapo}
\ee
with $\rho_m=\rho_{m,0}a^{-3}=3H_0^2\Omega_{m,0}a^{-3}$.
Using this expression in Eq.~(\ref{RDd2}) we find
\begin{equation}
\label{RDfR}
\frac{\Delta z}{\Delta t_0}(z)=\frac{\dot a_0- \dot a(z)}{a(z)}
=H_0\left\{(z+1)-\frac{1}{6}\frac{f(R)-Rf'(R)+6H^2f'(R)-2\rho_m}{H_0\dot R(t) f''(R)}\right\},
\end{equation} 
where $f$, $R$, $H$, and $\rho_m$ are functions of $z$ 
\footnote{Note that the presence of $f''$ in the denominator of this expression may lead to divergencies, since $f''$ is bound to be small if the theory is to yield an expansion close to that in GR +$\Lambda$ today. In the examples analyzed below, we checked that the product $H_0\dot R(t) f''(R)$ does not cause any divergencies.}.
We use 
$\frac{{\rm d}f}{{\rm d}z}=\frac{{\rm d}f}{{\rm d}R}\frac{{\rm d}R}{{\rm d}t}\frac{{\rm d}t}{{\rm d}z} $ 
and analogous expressions for other quantities to expand Eq.~(\ref{RDfR}) in powers of $z$. 
To first order\footnote{The second order term involves derivatives of the scale factor higher than 
the fourth, denoted by ${\ell}_0$.}, the result for an arbitrary $f(R)$ is given by
\bea
\label{RDz1f1}
\frac{\Delta z}{\Delta t_0}(z)&=&\left\{H_0+\frac{1}{6H_0f''_0}\left[\frac{1}{\dot{R}_0}\left(f''_0\dot{R}_0(R_0-6H_0^2)+2\dot{\rho}_{m,0}-12H_0\dot{H}_0f'_0\right) \right. \right. \qquad \qquad \qquad \qquad \qquad \qquad \qquad \nonumber \\
&& \qquad \qquad \qquad \qquad \qquad \qquad \qquad  \left. \left. + \left(\frac{\ddot{R}_0}{\dot{R}^2_0}+\frac{f'''_0}{f''_0}\right)\left(f_0+f'_0(6H_0^2-R_0)-2\rho_{m,0}\right)\right] \right\}z + {\cal O}(z^2).\  
\eea
Lastly, using that $R_0=6(\dot{H}_0 + 2H_0^2)$, $\dot{H}_0 = -H_0^2(1+q_0)$ and $\ddot{H}_0 = H_0^3(j_0+3q_0+2)$ together with the definitions of the kinematical parameters, we obtain
\bea
\label{RDz1f2} 
\frac{\Delta z}{\Delta t_0}(z)&=&\left\{ \left[6q_0 H_0^2 f'_0 +f_0 -6\Omega_{m,0} H_0^2\right](j_0-q_0-2)^2 \frac{f'''_0}{6f''_0} 
 +\right. \nonumber \\
&& \quad + H_0^2f''_0\left[(j_0-2)^2-q_0(3q_0+(q_0-j_0)^2-2j_0)\right]+ f'_0[q_0(q_0^2+6q_0+2j_0+s_0)+2j_0+4]\nonumber \\
&&\left. \qquad \quad +\frac{f_0}{36H_0^2}(-s_0-q_0^2-6-8q_0)
+\frac{\Omega_{m,0}}{6}(s_0+3j_0+5q_0+q_0^2)
 \right\}  \frac{z}{f''_0H_0(j_0-q_0-2)^2} + {\cal O}(z^2).\ 
\eea   
The dependence of the RD with the given theory is manifest in Eq.~(\ref{RDz1f2}) through $f$ and its derivatives evaluated today. 
We can now compare the linear term in $z$ of the kinematical and dynamic approaches to the RD, given by Eqs.~(\ref{RDz3}) and (\ref{RDz1f2}), respectively. 
The result is  
a relation between $f(R)$, its derivatives and the kinematical parameters, all evaluated at $t = $ today:
\bea
\label{restrictionf} 
\left\{[q_0(q_0(q_0+6)+2j_0+s_0)+2(j_0-2)]f'_0 +[s_0+q_0(q_0+8)+6]f_0-[q_0(q_0+5)+s_0+3j_0]\Omega_{m,0}\right\}f''_0 + & &\nonumber \\
\left[(q_0-j_0)^2+4(1+q_0-j_0)\right]\left\{f_0f'''_0+6H_0^2[(q_0f'_0-\Omega_{m,0})f'''_0 +(f''_0)^2]\right\}&=&0.\ \qquad
\eea 
Notice that the restriction to the first order in $z$ is not related to actual measurements of the RD for sources with $z\ll 1$, but to the fact that 
the second order term depends on ${\ell}_0$, 
for which there are no observational limits available. 
Note also that Eq.~(\ref{restrictionf}) is a necessary condition for any $f(R)$ theory to describe the variation of the RD with $z$. By equating higher orders of $z$  from Eqs.~(\ref{RDz3}) and (\ref{RDz1f2}) we would obtain more (actually, an infinite number of) necessary conditions on $f(R)$ and its derivatives. If the theory under discussion is to describe the RD at all orders in $z$, all these conditions should be satisfied. 

We shall see next how Eq.~(\ref{restrictionf}) constrains the value of the parameters of a given $f(R)$, by applying it to two examples.

\subsection{Example 1: The Theory of Hu and Sawicki}
\label{hs}
Let us start with the theory introduced by \citet{HuSawicki2007}, which is given by
\begin{equation}
f(R)=R -m^2\frac{c_1(R/m^2)^n}{c_2(R/m^2)^n +1},
\end{equation}
where $n>0$,  $c_1$ and $c_2$ are dimensionless parameters and the mass scale is $m^2\equiv \kappa^2\bar \rho_0/3$, with $\bar \rho_0$ the average density today.

\begin{figure}[ht!]
\begin{center}
\resizebox{100mm}{!}{\includegraphics[angle=-90]{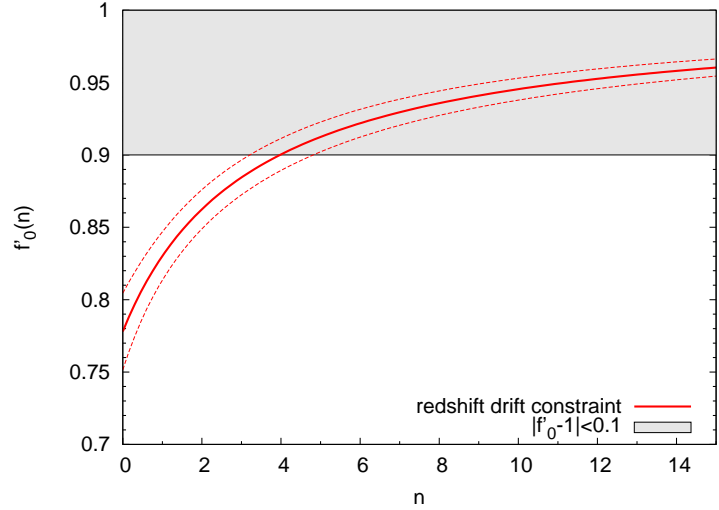}}
\caption{Constraints on the parameter space of the theory by Hu and Sawicki.  The red line corresponds to the relation between $f'_0$ and $n$ that follows  from Eq.~(\ref{restrictionf}). The dashed lines represent the error propagation arising from the kinematical parameters and the density matter at the present epoch. The shady region indicates the values of $f'_0$ in agreement with solar system tests.}
\label{plot_fR_HS} 
\end{center}
\end{figure}

For values of the curvature high compared with $m^2$
(which is actually the case if the current accelerated expansion 
is to be not very different today from 
that in GR+$\Lambda$, see \citet{HuSawicki2007}), $f(R)$ may be expanded as
\be
\label{high_curv}
\lim_{m^2/R\rightarrow 0} f(R) \simeq R -\frac{c_1}{c_2}m^2+\frac{c_1}{c_2^2}m^2\left(\frac{m^2}{R}\right)^n
\ee
and, at finite $c_1/c_2^2$, the theory can approximate the expansion history of the $\Lambda$CDM model \citep{HuSawicki2007}.  In this regime, the parameters $c_1$ and $c_2$ must satisfy the relation 
\be
\label{c1/c2}
\frac{c_1}{c_2}\approx6\frac{\tilde{\Omega}_{\Lambda}}{\tilde{\Omega}_{m}}.
\ee 
For the flat $\Lambda$CDM expansion history, Eq.~(\ref{high_curv}) yields 
\bea
R&\simeq& 3m^2\left(a^{-3}+4\frac{{\tilde{\Omega}}_{\Lambda}}{{\tilde{\Omega}}_{m}}\right), \\
f'&\simeq& 1-n\frac{c_1}{c_2^2}\left(\frac{m^2}{R}\right)^{n+1},
\eea
and at the present  epoch,
\bea
\label{R0}
R_0 & \simeq & m^2 \left(\frac{12}{{\tilde{\Omega}}_{m,0}}-9\right), \\
\label{fR0}
f'_{0} & \simeq & 1 -n\frac{c_1}{c_2^2}\left(\frac{12}{{\tilde{\Omega}}_{m,0}}-9\right)^{-n-1}. 
\eea 
Using Eq.~(\ref{fR0}), $c_1/c_2^2$ can be expressed in terms of $n$, $f'_0$ and ${\tilde{\Omega}}_{m,0}$. In addition, higher order derivatives of $f$ can also be written in terms of the same quantities. 
Hence from now on we set ${\tilde{\Omega}}_{m,0} = 0.274 \pm 0.007$ \citep{Komatsu2011} and leave $f'_0$ and $n$ the only free parameters of the  theory. With these considerations, Eq.~(\ref{restrictionf}) yields 
\bea
\label{constraint_HS}
f'_{0}(n)&=&\frac{1}{A}\left\{[q_0(5q_0-6j_0+8+(q_0-j_0)^2)+(j_0-2)^2]4n
+[j_0(j_0-8)-2q_0(j_0+q_0+8)-3(s_0+2)]6{\tilde{\Omega}}_{m,0} \right. \qquad \quad \nonumber \\
&& \left. \qquad \qquad \quad \ +[2(j_0+1)+q_0(q_0+6)+s_0]9{\tilde{\Omega}}_{m,0}^2
+[q_0(2(10+3q_0-2j_0)+(q_0-j_0)^2)+2(s_0+6)]4\right\},\ 
\eea
with
\bea
A&=&[q_0((q_0-j_0)^2-6j_0+8+5q_0)+(j_0-2)^2]4n-[3s_0+18+22q_0-(j_0-2)^2+(2q_0+1)^2)]6{\tilde{\Omega}}_{m,0} \qquad \qquad \quad \nonumber \\
&&\qquad \qquad \qquad \qquad +[s_0+q_0^2+8q_0+6]9{\tilde{\Omega}}_{m,0}^2 +[q_0(2(5q_0+16-4j_0)+(q_0-j_0)^2)+2(s_0+10-2j_0)]4.
\eea
Expression (\ref{constraint_HS}) gives a relation between the parameters $f'_0$ and $n$ in terms of ${\tilde{\Omega}}_{m}$ and the kinematical parameters $H$, $q$, $j$,  and $s$, all evaluated today.  
We shall take the values 
$q_0 = -0.669 \pm 0.052$, $j_0=0.284 \pm 0.151$ and $s_0=-0.680 \pm 0.456$ \citep{Capozziello2011b}.
In Figure \ref{plot_fR_HS} we plot the relation $f'_0(n)$ provided by our cosmographical approach to the RD
 combined with the  expansion of the expression of $H(z)$ for $f(R)$ theories
\footnote{
Due to the current observational limitations to measure accurately the kinematical parameters, an appropriate error propagation treatment was 
applied in the analysis.}. The curve tends asymptotically to $f'_0 = 1$, which corresponds to the GR limit. 
The plot also displays 
the limit obtained from solar system tests, given by $|f'_{0}-1|<0.1$ \citep{HuSawicki2007}. 
We find that actually $1-f'_0<0.1$ and, from this limit, values for $n$ larger than approximately $3$ are favoured, thus discarding low values for $n$, and allowing for large values, in accordance with the findings 
of 
\citep{Martinelli2012b}.  


\subsection{Example 2: Exponential Gravity}
\label{lin}
\begin{figure}[ht!]
\begin{center}
\resizebox{100mm}{!}{\includegraphics[angle=-90]{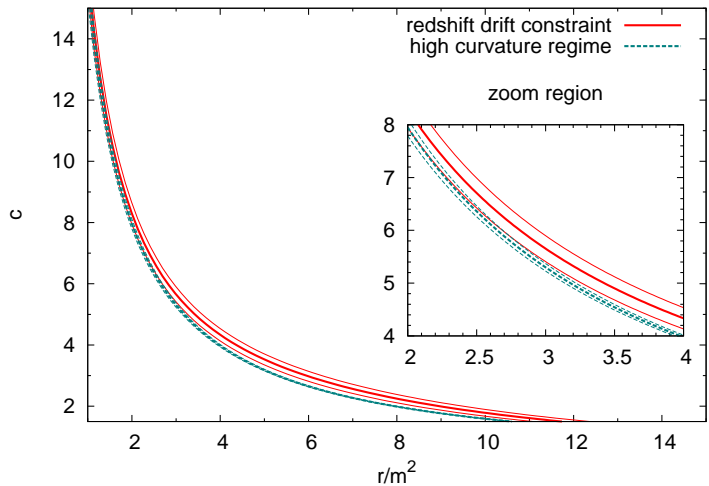}}
\caption{Constraints coming from our approach to the redshift drift (continuous line) and the relation (\ref{cr}) between the parameters of the theory 
(dashed line). In both cases, the thin lines represent the error propagation arising from the  kinematical parameters and the matter density at the present epoch.}
\label{plot_Linder1}
\end{center}
\end{figure}

\noindent Next we shall analyze the restrictions that follow  from Eq.~(\ref{restrictionf})
on the 
choice of $f(R)$ proposed by \citet{Linder2009}: 
\begin{equation}
\label{fRLinder}
f(R)=R - cr  (1-\exp(-R/r)),
\end{equation}
where $c$ and $r$  are two (positive) parameters of the model.  
This $f(R)$ was specifically designed to (i) 
avoid the inclusion of an  implicit cosmological constant (since it vanishes in the low curvature limit), (ii) reduce to GR for high values of the curvature,
(iii) incorporate a transition scale (given by $r$) to be fitted from observations (instead of set equal to $R_0$), and
(iv) restore GR for locally high curvature systems
such as the solar system or galaxies.  
As shown in \citep{Linder2009}, 
the product $cr$ is given in terms of $\Omega_{m,0}$ by 
\begin{equation}
\label{cr}
cr=6m^2(\Omega_{m,0}^{-1}-1).
\end{equation}   
The use of Eq.~(\ref{restrictionf}) for the current choice of $f(R)$
yields a relation between the dimensionless parameters $c$ and $r/m^2$ given by
\bea
c\left(\frac{r}{m^2}\right) &=& \frac 1 B \left\{
\left[(3+s_0+2j_0(q_0+1)-(q_0-1)^2)-(q_0(q_0+5)+s_0+q_0^2+3j_0)\Omega_{m,0}\right]6\Omega_{m,0}\frac{r}{m^2} \right. \qquad \qquad \qquad \qquad \nonumber \\
&& \qquad \qquad \qquad \qquad \qquad \qquad \qquad \qquad\left. +[4(1+q_0-j_0)+(q_0-j_0)^2]36\left(\Omega_{m,0}-1\right)
\right\}\exp\left(\frac{6(1-q_0)}{\frac{r}{m^2}\Omega_{m,0}}\right), \ \ \ 
\eea
%
with
\bea
B& =&\left\{[q_0(q_0+8)+6+s_0]\Omega_{m,0}\left(\frac{r}{m^2}\right)\left[(\exp\left(\frac{6(1-q_0)}{\frac{r}{m^2}\Omega_{m,0}}\right)-1\right]
-6[4(1+q_0-j_0)+(q_0-j_0)^2]\exp\left(\frac{6(1-q_0)}{\frac{r}{m^2}\Omega_{m,0}}\right) \right.\nonumber \\
&& \left. +6[q_0(q_0(q_0+7)+4+s_0)+(j_0-1)^2-1] \right\}\Omega_{m,0}\left(\frac{r}{m^2}\right)
-[(j_0-2)^2+q_0((q_0-j_0)^2+8-6j_0+5q_0)]36, 
\eea
which is plotted along with Eq.~(\ref{cr}) in Figure \ref{plot_Linder1}
\footnote{We have taken into account in the plot that 
the distance to the cosmic microwave background
last scattering surface in this theory agrees with the $\Lambda$CDM model with
the same present matter density to 0.2\% if 
$c\geq 1.5$.}.
The strips formed by both curves and the corresponding errors juxtapose 
for all values of 
$c \gtrsim 6 $, which imply that $r/m^2 \lesssim 2.7$.      
Notice that the range of possible values for the parameter $c$ that follows from our method improves the previous bound ($c\geq 1.27$) obtained in \cite{Yang2010}.

\section{Discussion}
\label{sec:discussion} 
 
We have presented a method to set constraints on the parameters of $f(R)$  theories of
gravitation. It is based on the comparison of two series expansions of any observable that depends on $H(z)$. The first expansion is of the cosmographical type ({\em i.e.} independent of the dynamics of the theory), while the second uses the dependence of $H$ with $z$ furnished by any $f(R)$. The comparison of the two expansions yields relations between $f$, its derivatives, and the kinematical parameters, all evaluated at $z=0$. These relations must be satisfied by any $f(R)$.  
We showed that when the observable is the redshift drift, the method 
yielded limits on the $n$ parameter of the $f(R)$ introduced in \citet{HuSawicki2007} that are in agreement with previous findings (obtained without using the redshift drift). In the case of the exponential gravity theory introduced by \citet{Linder2009}, the bound we obtained in the parameter $c$ is stronger than previously obtained limits. We also presented for the first time a bound on the parameter ($r/m^2$). As a byproduct, the cosmographic expression for the redshift drift given 
in Eq.~(\ref{RDz3}) was obtained, that must be obeyed by any theory. 
It is worthwhile noting that the
method we introduced is not restricted to $f(R)$theories: except for algebraic 
problems in particular examples, it can be applied to any alternative
theory of gravity under the assumption of homogeneity and isotropy.

To close, we would like to emphasize that the bounds obtained by the method developed here can be analyzed toghether with those coming from the observations mentioned in the Introduction as well as 
other means (such as energy conditions \citep{Santiago2006}), with the aim of deciding whether a given $f(R)$ theory 
is consistent with the available data.

\section*{Acknowledgements}

FATP acknowledges support from CONICET and the Programa de Doctorado Cooperativo CLAF/ICTP, and the hospitality of UERJ and CBPF. SEPB would like to acknowledge support from FAPERJ, UERJ, and ICRANet-Pescara. 


\bibliographystyle{apsrev} 
\bibliography{bibliography} 

\end{document}